\documentclass[letter]{aa}
\usepackage{graphicx}
\usepackage{natbib}
\newcommand{\be}{\begin{eqnarray}}
\newcommand{\ee}{\end{eqnarray}}

\newcommand {\nbodypp}{\textsc{\mbox{nbody6\raise.4ex\hbox{\tiny++}}}}

\newcommand {\Msun} {\mbox{M$_{\odot}$}}

\newcommand {\Zsun} {\mbox{Z$_{\odot}$} }

\begin{document}

\title{How Universal are the Young Cluster Sequences?\\ - the Cases of LMC,
SMC, M83 and the Antennae}
\author{S. Pfalzner$^1$, A. Eckart$^{1,2}$} 
\institute{Physikalisches Institut, University of Cologne, Z\"ulpicher Str. 77, 50937 Cologne, Germany
\and Max-Planck-Institut f\"ur Radioastronomie, Auf dem H\"ugel 69, 53121 Bonn, Germany}
\date{}
\titlerunning{How Universal are the Young Cluster Sequences?}

\abstract
{}
{Recently a new analysis of cluster observations in the Milky Way found evidence that clustered star 
formation may work under tight constraints with respect to cluster size and density, implying the presence
of just two sequences of young massive cluster. These two types of clusters each expand at
different rates with cluster age.}
{Here we  investigate whether similar sequences exist in other nearby galaxies.} 
{We find that
while for the extragalactic young stellar clusters the overall trend in the cluster-density scaling is 
quite comparable to the relation obtained for Galactic clusters, there are also possible difference. 
For the LMC and SMC clusters the densities are below the Galactic data points and/or the core radii are 
smaller than those of data points with comparable density. For M83 and the
Antenna clusters the core radii are possibly comparable to the Galactic clusters but it is not clear whether
they exhibit similar expansion speeds. These findings should serve as an incentive to perform more 
systematic observations and analysis to answer the question of a possible similarity between
young galactic and extragalactic star clusters sequences.}
{}

\maketitle

\section{Introduction}

\begin{figure}[t]
\resizebox{\hsize}{!}{\includegraphics[angle=-90]{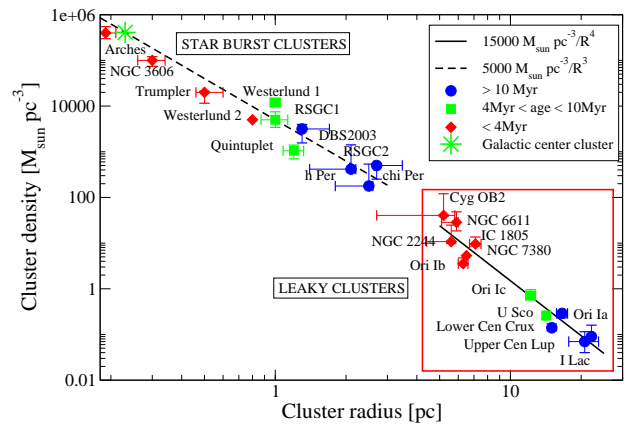}}
\caption{{Cluster density as a function of cluster size for clusters more massive than 10$^3$ \Msun
as published by Pfalzner (2009).
references therein.}}
\label{fig:cluster_rad}
\end{figure}

Lada \& Lada (2003) showed that in our Galaxy most stars do not form in isolation but in cluster environments.
These young clusters typically consist of thousand stars or more with densities ranging 
from $<$0.01 to several 10$^5$ \Msun pc$^{-3}$ 
The wide variety of observed cluster densities lead 
to the assumption that clusters are formed over this 
entire density range. Albeit Maiz-Apellaniz (2001) noticed the existance of two types of clusters
in the Galaxy, only recently Pfalzner (2009) found that massive clusters develop 
in a bimodal way as in Fig.\ref{fig:cluster_rad}, where the red, green and blue symbols represent 
clusters with ages $t_c <$ 4 Myr,  4 Myr $ < t_c <$ 10 Myr,  10 Myr $< t_c <$ 20 Myr, respectively.
Two well-defined sequences in the density-radius plane emerge showing the {\em bi-modal nature of the
cluster evolution.} 

\begin{figure}[t]
\resizebox{\hsize}{!}{\includegraphics[angle=0]{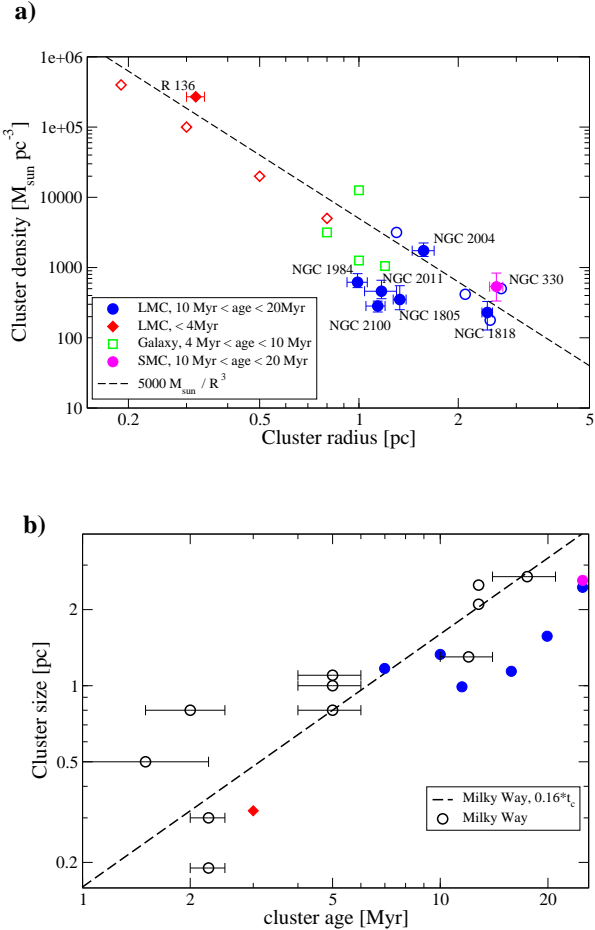}}
\caption{a) Cluster density as a function of cluster radius and b) cluster size as a function of cluster age. Here the LMC and SMC (purple circle) clusters are shown as filled
symbols and the Milky Way as open symbols (same clusters as in Fig.~\ref{fig:cluster_rad}).
The values for the clusters in the LMC are taken from Mackey \& Gilmore 2003a and McLaughlin \& van der Marel 2005, 
the SMC value from Mackey \& Gilmore 2003b.}
\label{fig:LMC_cluster_size_age}
\end{figure}

Pfalzner (2009) classified the two modes as starburst and leaky cluster sequences.
The starburst cluster sequence implies a population of compact clusters (0.1 pc) with high initial densities 
(10$^5$ - 10$^6$ \Msun pc$^{-3}$) which then expand with the mass-density decreasing as $\sim R^{-3}$ (an actual $2\sigma$-fit
gives a $R^{\alpha}$-dependence with $\alpha$=-2.71 $ \pm $ 0.32) and evolve to have sizes of a few pc over a period of 10 Myr or longer.
Prominent members of this type of cluster are for example Arches, NGC 3603 and Westerlund 1. 
The leaky cluster sequence implies the creation of a second population of diffuse clusters ($\sim$ 5pc) which
expand loosing mass during the process, until they have sizes of a few tens of pc.
The expansion also follows a predefined track, but here the density decreases as $\sim R^{-4}$ 
(actual fit value $\alpha=-$4.07$\pm0.33$). Here NGC 6611, Ori 1a-c or U Sco are typical examples.
Note, the cluster radii were not determined exactly the same way for all
clusters shown in Fig.~\ref{fig:cluster_rad}. However, 
star burst cluster values all represent the core radii 
apart from $\chi$ Per and h Per (for a discussion on the determination of these two radii see  Pfalzner 2009).

It follows that star formation occurs only under an extremely limited set  of conditions, and may require a 
fundamental revision of star formation hypotheses in our Galaxy.  This immediately raises the questions of the origin 
of these two distinct cluster sequences and whether similar density-radius correlations could be found in other 
galaxies. This Letter addresses the latter question.


Most extragalactic clusters one observes are likely to belong to the starburst cluster sequence because although their extension is on average
smaller the luminosity of their high number of O-stars is much easier to detect than the smaller number of O stars spreading 
over a larger area in leaky clusters. So in all figures subsequent to Fig.~1 only the starburst clusters of the Galaxy
are shown for comparison and the radial extent is the core radius. 

\begin{table*}
\begin{center}
\begin{tabular}{l|*{5}{l}}
\hline
\\[-2ex]
Galaxy & radius determination   &  mass determination &  background  &  age determination\\\hline
\\[-2ex]
LMC$^1$          &  King profile fit        & from mass/light ratio from    & mean surface brightness     & combination of IMF $^8$ \\
             &  half-surface brightness & evolutionary code$^5$         & in annulus at r $\gg R_c$   & and evolutionary code$^5$  \\
SMC$^2$          &  King profile fit        & from mass/light ratio from    & mean surface brightness     & Based on color-magnitude  \\
             &  half-surface brightness & evolutionary code$^5$        & in annulus at r $\gg R_c$    & diagram$^{12}$  \\
Antennae$^3$     &  Fit with ISHAPE code$^6$    & UBVI-colors and mass/light   &   Fit with ISHAPE code      & LICK line strength indices  \\
             &                          & ratio$^7$ with extinction-   &                             & and $^{11}$  \\
             &                          & corrected SSP   &                             &                        \\
M 83$^4$         &  Fit with ISHAPE code$^6$    & Isochromes$^9$  &   Fit with ISHAPE code      & UVBI-colors, 2-color and  \\
             &                          &                               &                          & S-sequnce diagram $^{10}$
\end{tabular}
\caption{Determination methods of radius, mass and age and background consideration ($^1$Maxkey \& Gilmore 2003a, 
$^2$ Mackey \& Gilmore 2003b, $^3$ Mengel et al. 2008, $^4$ Larsen \& Richter 2004, $^5$Fioc \& Rocca-Volmerangen 1997, $^6$ Larsen 1999, 
$^7$Anders \& Fritz-von-Alvensleben 2003, $^8$ Kroupa et al. 1993,  $^9$ Girardi 2000, $^{10}$ Girardi 1995, $^11$ Schweizer 2004 $^{12}$ Chiosi 1995.
\label{table:method}}
\end{center}
\end{table*}

Currently the Milky Way is not in a intense star formation phase accounting for the 
scarcity of starburst clusters in the Galaxy. Another reason is that extinction at low 
latitudes hampers the detection of distant Galactic clusters. By contrast, there are 
starburst and interacting galaxies such as the "Antennae", where many clusters 
with masses $>$ 10$^4$ \Msun\ are observed with ages $<$1Gyr (Zhang \& Fall 1999). 
In between, there exist the dwarf starburst galaxies such as NGC 4214, where several 
massive clusters with ages $<$100 Myr are visible in the central regions.

This Letter shows a first investigation of whether similar development tracks  
can be found for clusters in such different galaxies.
Observationally one is restricted to nearby galaxies, and even there radii are only resolved for a 
very limited sample.
Here we scanned the literature for clusters younger than 30 Myr in the Large Magellanic Cloud (LMC), 
Small Magellanic Cloud (SMC), Messier 83 (M83) and the Antennae.


\section{Limitations}

Comparing clusters in the Milky Way with extragalactic clusters intrinsically has its limitations. Especially
age dating differs in both cases - whereas for Milky Way clusters ages can be based on individual 
stars, in distant clusters they are determined from integrated colors. Even more complex is the comparison
of radii as there exist different definitions resulting in various measurement methds. In addition,
effects of the different spatial resolutions, background and used photometric bandpasses can strongly influence
the results. In general we compare here the core radii of the different clusters (Note Fig.1 contains partly
different definitions of radii, for a discussion see Pfalzner 2009). Ideally one would use the same age, mass and radii 
determination techniques for all investigated clusters. However, unfortunately for most of the clusters no public data are 
available, so that in the following literature data were used. Table 1 lists for each of the investigated
galaxies the methods used for age, mass, radius and background determination.

\section{Magellanic Clouds}

Since the LMC and SMC are at distances of ≃ 50 kpc and 60 kpc \cite{schaefer:08}, respectively, their
system of star clusters can be studied in much more detail than those of more distant hosts.
Compared to the Galaxy, the Magellanic Clouds are gas rich and metal poor, thus
providing a different environment where stars form and evolve.

\subsection{LMC}

Rich clusters in the LMC span a wide range in ages, 6 $\leq \log (yrs) \leq 10$. Most clusters younger than
$\simeq$ 3 Gyr  have metalicities between 1/2 and 1/3 of the solar values \cite{santiago:08}.
Although thousands of clusters are known in the LMC, only a few fulfill the specific conditions of our study, i.e.
cluster age $<$  20 Myr with known mass and radial extent. Fig.~\ref{fig:LMC_cluster_size_age}a) shows the density
as function of cluster radius for LMC clusters. The data are taken from 
Maykey \& Gilmore (2003) and compared to the starburst cluster sequence of the Galaxy.
Most of the LMC cluster data points lay below those of the Galaxy. The same applies comparing the cluster age to
cluster radius (see Fig.~\ref{fig:LMC_cluster_size_age}b).

\subsection{SMC}

The Small Magellanic Cloud (SMC) is the closest star forming dwarf galaxy ($\sim$ 60 kpc). Its present 
day metalicity (Z=0.004) and low dust content (30 times smaller than in the Milky Way) make the SMC an
prototype late type dwarf. At such low metalicity one expects that the resulting reduced
stellar wind can modify the early evolution of clusters as the powerful winds characteristic of systems
of solar metalicities do not exist and therefore might not be able to remove the gas left from star formation.
Unfortunately we found only one cluster younger than 30 Myr with known density and radius. It lies a bit above 
the Galactic star cluster track (see Fig.~\ref{fig:LMC_cluster_size_age})a.


\subsection{Discussion}

The very limited data available for the LMC and SMC give a first indication of a similar sequential track for 
the LMC and SMC as in the Galaxy possibly at lower density values for the LMC. 
One possible reason for such a lower track could be 
the lower metalicities in the LMC system (Cepheid metalicities of Z=0.0091+/-0.0007 and 0.0050+/-0.0005
for the LMC and SMC, respectively, compared to Z=\Zsun$\sim$0.02;
Keller \& Wood, 2006).
The lower metalicities may result in a larger Jeans mass and therefore
larger initial mass limit in general (Clark \& Bromm 2002).
At the same mass lower metalicity stars are more luminous
(e.g. Bertelli et al. 2008). Hence, including the effect of mass
segregation, the centers of low metalicity star clusters may
be overluminous  compared to higher metalicity clusters,
resulting underestimating their sizes. However, It is unclear
whether this effect is strong enough to result in the observed difference in the
size measurements.

The uncertainties in the distance modulus of the LMC 
(LMC: 18.54+/-0.018; SMC: 18.93+/-0.024; Keller \& Wood, 2006)
is small and cannot be responsible for the entire size discrepancy.
The clusters are seen against the stellar background of the LMC/SMC.
This may result in an underestimation of the contribution of
the more extended off center stellar cluster
brightness distribution that is not dominated by luminous stars.
This effect will lead to an underestimation of the cluster size.

\section{The Antennae}

The Antennae are the nearest and best-studied pair of merging galaxies, consisting of two large 
spirals that began to collide a few × 10$^8$ yr ago. The ongoing merger is almost certainly 
responsible for the large population of clusters. Understanding the formation and disruption 
of clusters in this setting is important because it represents a latter-day example of the 
hierarchical formation of galaxies, a process that operated even more effectively in the early 
universe. Fall et al. (2005) find a steep decline in the number of cluster as a  function of age and 
interpret as a sign of intensive  disruption of the clusters. They conclude that the short 
timescale on which the clusters are disrupted indicates that most of them are not gravitationally 
bound. So the Antennae seem to show a high rate of ``infant mortality'' similar to  the Milky Way.
As the clusters are initially disrupted mainly by internal processes, they expect the 
age distribution to be largely independent of the properties of the host galaxy. 

\begin{figure}[t]
\resizebox{\hsize}{!}{\includegraphics[angle=-90]{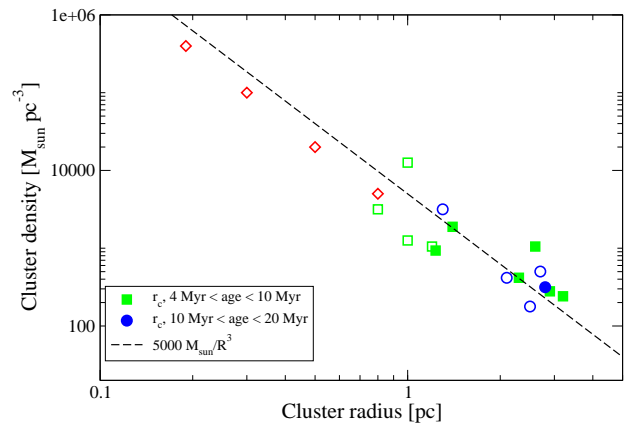}}
\caption{Cluster size as a function of cluster age. The Antennae clusters  (data from \cite{bastian:09}) are shown as filled symbols. The Milky Way clusters are indicated in addition (open symbols).
}
\label{fig:cluster_size_age_ant}
\end{figure}

As only values of the effective radii $r_{eff}$ (half-light radii) are given for the
Antennae clusters in Fig.~\ref{fig:cluster_size_age_ant} core radii $r _c$  were derived  
assuming a factor of 3 difference between $r_{eff}$ and $r_c$ as deduced 
for globular clusters\cite{schweizer:04}. Whether this relation holds as well for young clusters is
uncertain. The so deduced core radii fit  quite well onto the Galactic starburst cluster sequence,
but in general the ages of the clusters are younger than that of equivalent Galactic starburst clusters.

\section{M83}

\begin{figure}[t]
\resizebox{\hsize}{!}{\includegraphics[angle=-90]{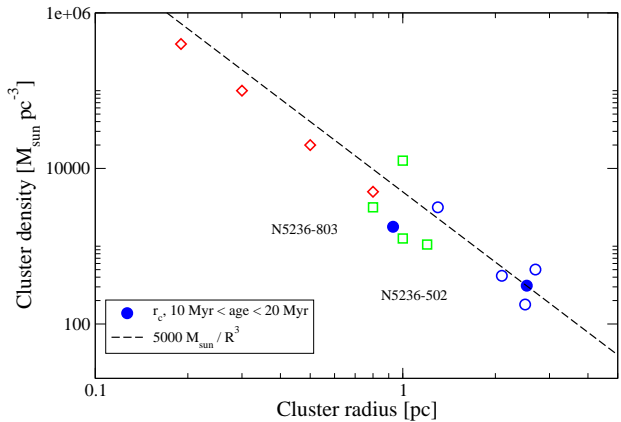}}
\caption{
Cluster size as a function of cluster age. The M83 clusters \cite{larsen:04} are shown as filled
symbols. The Milky Way clusters are indicated in addition (open symbols).}
\label{fig:cluster_size_age_m83}
\end{figure}

The nearby spiral galaxy M83 is known to have a rich population of 
relatively young clusters, most likely due to its high star formation rate.
Using the cluster data by Larsen \& Richtler (2004) Fig.~\ref{fig:cluster_size_age_m83} shows the cluster density 
as a function of the effective radius. Like for the Antennae galaxy we deduced the core radii by
assuming a factor of 3 difference between $r_{eff}$ and $r_c$ as deduced 
for globular clusters\cite{schweizer:04}. The so obtained data lie on the galactic starburst cluster sequence.

\section{Conclusion}

For the extragalactic young stellar clusters the overall trend in the cluster radius versus cluster density
relation seems to fit the relation obtained for Galactic starburst clusters by Pfalzner (2009). A possible difference
is indicated for the LMC clusters, where the densities are mostly below the comparable 
Galactic data at equal radius, or the radii are smaller than those of data points with 
comparable density. This might indicates that the sizes for LMC clusters are systematically smaller than 
the Galactic cluster sizes. This could be explained by intrinsic observational effects.


However, the sample size of extragalactic young clusters ($<$ 20 Myr ) with known radii and masses is so small that this
investigation can only be regarded as a first hint that the young starburst cluster sequence might also exist in galaxies other than
the Milky way. In this study only literature values of age, mass and radius were used mostly 
obtained by different methods. Further studies with a larger sample size and more homogenous data acquisition will be 
required to confirm this result. 
               

\bibliographystyle{apj}

\begin{thebibliography}{}

\bibitem[Anders 2003]{anders:09}
Anders, P., Fritz-von-Alvensleben, U. 2003 A\&A 401, 1063.


\bibitem[Bastian et al. \- 2009]{bastian:09}
Bastian, N., Trancho, G., Konstantopoulos, I., Miller, B. 2009, astro-ph 0906.2210.


\bibitem[Bertelli et al. \- 2008]{bertelli:08}
Bertelli, G., Girardi, L., Marigo, P., Nasi, E. 2008, A\&A 484, 815.


\bibitem[Bromm \& Clarke 2002]{clark:02}
Bromm, V. \& Clarke, C. 2002 ApJ 566, L1.

\bibitem[Chiosi 1995]{chiosi:95}
Chiosi, C.  Vallenari, A., Bressan, A., Deng, L., Ortolani, S. 1995 A\&A 293, 710.


\bibitem[Fall et al. \- 2005]{fall:05}
Fall, S. M., Chandar, R., Whitmore, B. C., 2005, ApJ 631, L133.


\bibitem[Fioc \& Rocca-Volmerange 1997]{fioc:97}
Fioc, M. \& Rocca-Volmerange, B. 1997, A\&A, 326, 950.

\bibitem[Gieles \& Bastian 2008]{gieles:08}
Gieles, M., Bastian, N., 2008, A\&A 482, 165.

\bibitem[Girardi et al. 1995]{girardi:95}
Girardi, L., Chiosi, C., Bertelli, G.A. 1995, A\&A 298, 87.

\bibitem[Girardi et al. 2000]{girardi:00}
Girardi, Bressan, A., Bertelli, G.A., Chiosi, C. 2000, A\&AS 141, 371.

\bibitem[Keller \& Wood 2006]{keller:06}
Keller, S. C. \& Wood, P. R., 2006, ApJ 642, 834.

\bibitem[Kroupa et al. 1993]{kroupa:93}
Kroupa, P., Tout, C.~A. \& Gilmore, G.~F. 1993, MNRAS 262, 545.


\bibitem[Larsen \& Richtler 1999]{larsen:99}
Larsen, S. S. \& Richtler, T. 1999, A\&A 345, 59.

\bibitem[Larsen \& Richtler 2004]{larsen:04}
Larsen, S. S. \& Richtler, T. 2004, A\&A 427, 495.

\bibitem[Mackey \& Gilmore 2003a]{mackey:03}
Mackey, A.D. \& Gilmore, G.F.  2003, \mnras, 338, 85.

\bibitem[Mackey \& Gilmore 2003b]{mackey:03b}
Mackey, A.D. \& Gilmore, G.F.  2003, \mnras, 338, 120.

\bibitem[Maiz-Apellaniz 2001]{maiz:01}
Maiz-Apellaniz, J., 2001, ApJ 563, 151.

\bibitem[McLaughlin \& van der Marel 2005]{mclaughlin:05}
McLaughlin \& van der Marel 2005 ApJS 161 304.

\bibitem[Mengel et al. 2008]{mengel:08}
Mengel, S., Lehnert, M.D., Thatte, N.A., Vacca, W.D., Whitmore, B., Chandar,  2008, \aa, 489, 1091.

\bibitem[Mengel et al. 2005]{mengel:05}
Mengel, S., Lehnert, M. D., Thatte, N., Genzel, R., 2005, A\&A 443, 41.


\bibitem[Pfalzner 2009]{pfalzner:09}
Pfalzner, S. 2009, A\&A 498, L37.

\bibitem[Santiago 2008]{santiago:08}
Santiago, B., In {\it The Magellanic System: Stars, gas and galaxies}, ed. van Loon, J. \& Oliveira, J., Proceed. IAU 256,
(2008)

\bibitem[Schaefer 2008]{schaefer:08}
Schaefer, B., 2008, AJ 135, 112.

\bibitem[Schweizer 2004]{schweizer:04}
Schweizer, F. 2004 in Proc. of {\it Formation and Evolution of Massive Young Star Clusters}
ASP Conference Ser., eds. H. Lamers, A. Nota \& l.J. Smith





\end{thebibliography}


\newpage
\onecolumn
\section{Online material}

\newpage

\begin{table}
\begin{center}
\begin{tabular}{l|*{8}{c}}
\hline
\\[-2ex]
Identifier & log(age)    &  $\Delta$ age &  radius   &  $\Delta$ radius & $M_{total}$ & $\Delta M_{total}$ & log$(\rho)$ & $\Delta \rho$\\
           & [Myr]        &               &  [pc]     &                  & [\Msun]     & &                  [\Msun/pc$^{-3}$] &          \\\hline
\\[-2ex]
NGC 1805$^1$ &   7.0   &  0.3  &  1.33  &  0.06 & 3.52 & 0.13 & 2.54 & 0.2\\
NGC 1818$^1$ &   7.4   &  0.3  &  2.45  &  0.09 & 4.13 & 0.15 & 2.35 & 0.2\\
NGC 1984$^1$ &   7.06  &  0.3  &  0.99  &  0.07 & 3.38 & 0.24 & 2.79 & 0.2\\
NGC 2004 $^1$&   7.30  &  0.2  &  1.57  &  0.13 & 4.43 & 0.38 & 2.32 & 0.2\\
NGC 2011$^1$ &   6.99  &  0.3  &  1.17  &  0.12 & 3.47 & 0.14 & 3.24 & 0.12\\
NGC 2100$^2$ &   7.20  &  0.3  &  1.14  &  0.14 & 4.43 & 0.13 & 3.05 & 0.2\\
R 136$^1$    &   6.48  &  0.18 &  0.32  &  0.02 & 4.55 & 0.21 & 4.47 & 0.08\\
\end{tabular}
\caption{Properties of clusters younger than $log(t_c)$=8.00 in the LMC from
$^1$Mackey \& Gilmore 2003a and $^2$McLaughlin \& van der Marel 2006.
\label{table:nacc1}}
\end{center}
\end{table}

\begin{table}
\begin{center}
\begin{tabular}{l|*{8}{c}}
\hline
\\[-2ex]
Identifier & log(age)    &  $\Delta$ age &  radius   &  $\Delta$ radius & $M_{total}$ & $\Delta M_{total}$ & log$(\rho)$ & $\Delta \rho$\\
           & [Myr]        &               &  [pc]     &                  & [\Msun]     & &                  [\Msun/pc$^{-3}$] &          \\\hline
\\[-2ex]
NGC 330 &   7.4   &  0.4 &  2.61  &  0.12 & 4.58 & 0.2 & 2.72 & 0.2\\
\end{tabular}
\caption{Properties of clusters younger than $log(t_c)$=8.00 in the SMC from
\cite{mackey:03b}
\label{table:nacc1}}
\end{center}
\end{table}

\begin{table}
\begin{center}
\begin{tabular}{l|*{6}{c}}
\hline
\\[-2ex]
Identifier &   log(age) [Myr]   &  radius [pc]  & log(mass) [\Msun]& log$(\rho)$ [\Msun/pc$^{-3}$] \\\hline
\\[-2ex]
T54   &   6.9 $\pm$  0.1 &  3.7  &  4.8 $\pm$ 0.3 & 2.49 & \\
T 270 & $<$ 6.8          &  9.3  &  5.4 $\pm$ 0.3 & 1.89 & \\
T 324 & 6.5 - 6.8        &  7.7  &  5.2 $\pm$ 0.3 & 2.54 & \\
T 343 & 6.5 - 6.8        &  8.8  &  5.4 $\pm$ 0.3 & 1.96 & \\
T 365 & 6.5 - 6.8        &  4.3  &  5.3 $\pm$ 0.3 & 2.80 & \\
T 367 & 6.5 -6.8         &  6.6  &  5.2 $\pm$ 0.3 & 2.14 & \\
\end{tabular}
\caption{Properties of clusters younger than $log(t_c)$=7.50 in the Antennae from
\cite{bastian:09}
\label{table:nacc1}}
\end{center}
\end{table}

\begin{table}
\begin{center}
\begin{tabular}{l|*{6}{c}}
\hline
\\[-2ex]
Identifier &   log(age) [Myr]   &  $r_{eff}$ [pc]  & log(mass) [\Msun]& log$(\rho)$ [\Msun/pc$^{-3}$] \\\hline
\\[-2ex]
N5236-502   &  8.0 $\pm$  0.1   &  7.6 $\pm$  1.1  &  5.15 $\pm$  0.83  &  2.8 $\pm$  1.0 $\times $ 10$^3$  \\
N5236-805   &  7.1 $\pm$  0.2   &  2.8 $\pm$  0.4  &  4.16 $\pm$  0.67  &  1.6 $\pm$  1.1 $\times $ 10$^4$  \\
\end{tabular}
\caption{Properties of clusters younger than $log(t_c)$=7.50 in M 83 from
\cite{larsen:04}
\label{table:nacc1}}
\end{center}
\end{table}

\end{document}